\documentclass[prl,reprint, nofootinbib,amsmath,amssymb, aps, floatfix]{revtex4-2}

\usepackage{amssymb,amsmath}
\usepackage{bm}
\usepackage{graphicx, xcolor}
\usepackage{braket}
\usepackage{hyperref}
\usepackage{mathrsfs}
\usepackage{subcaption}
\usepackage{comment}
\usepackage{physics}
\usepackage{svg}
\begin{document}

\title{
The endless journey towards the horizon of a quantum black hole}

\author{Victor Franken, Thomas G. Mertens, and Bruno de S. L. Torres}
\affiliation{Department of Physics and Astronomy, Ghent University, Krijgslaan 299, 9000 Gent, Belgium}

\begin{abstract}
It is a problem of great importance in quantum gravity whether quantum effects lead to nontrivial structure near black hole horizons, and if their interiors are emergent features of a large $N$ limit. We show that the strongly chaotic nature of black holes, captured by random matrix statistics of their spectrum, makes it so that a suitably gravitationally dressed infalling observer takes an infinite amount of proper time to reach the horizon. This makes the black hole interior operationally inaccessible to any outside observer, marking a sharp breakdown of the large $N$ causal structure of black holes in quantum gravity. To model the experience of an infalling observer, we study the response of an infalling particle detector coupled to a matter field in a 2d quantum black hole background, finding a peak in the detector's excitation probability as it enters an infinitely long non-perturbative journey close to the black hole horizon. 
\end{abstract}

\maketitle

\textbf{\textit{Introduction--}} Defining the interior of a black hole beyond the semiclassical regime has been a central theme in quantum gravity \cite{Maldacena:2001kr,Kraus:2002iv,
Almheiri:2013hfa,
Marolf:2013dba,Papadodimas:2012aq,
Papadodimas:2013wnh,
Papadodimas:2013jku,Grinberg:2020fdj,Leutheusser:2021frk,Hamilton:2006fh,Lewkowycz:2016ukf,Almheiri:2017fbd,Jafferis:2020ora,Gao:2021tzr,deBoer:2022zps,Leutheusser:2022bgi,Leutheusser:2021qhd}. In classical general relativity, the interior region appears as the analytic extension of the metric solution behind the event horizon. This mathematical procedure is physically relevant because the affine time for a null geodesic to reach the horizon is finite, so a sufficiently long-lived observer can see the interior in finite time. In quantum gravity, however, we expect non-perturbative effects to affect not only the black hole singularity but also the entire black hole interior. This can lead to drastic changes in how the approach to the black hole interior is experienced beyond the semiclassical regime.
With this motivation in mind, we advocate for an observer-based test made by a localized probe that attempts to access the black hole interior in quantum gravity:

\emph{A necessary operational condition for the existence of the black hole interior beyond the semiclassical regime is that the affine time it takes a null geodesic to cross the horizon, when promoted to a quantum operator, remains finite.}

While simple in appearance, this condition has important technical challenges. One crucial feature of theories of dynamical gravity is that the notion of ``local observables'' can only be defined relationally, as general diffeomorphism invariance precludes any absolute notion of the ``location'' of a point~\cite{Rovelli:1990ph, Giddings:2005id, Donnelly:2016rvo}.
This means that local operators become meaningful only when ``gravitationally dressed'', implying that the observable is described relative to some other feature of the system serving as a reference frame.
Moreover, at the most fundamental level, all reference systems---i.e., the clocks and rods used to build a physical reference frame in the real world---are subject to the rules of quantum theory~\cite{Giacomini:2017zju}. Observables such as proper distances should thus be viewed as quantum operators, and nontrivial quantum-gravitational effects can emerge once the reference system's own degrees of freedom are treated quantum-mechanically.

\begin{figure}[t]
    \begin{subfigure}[t] {0.48\linewidth}  
\includegraphics[width=0.7\linewidth]{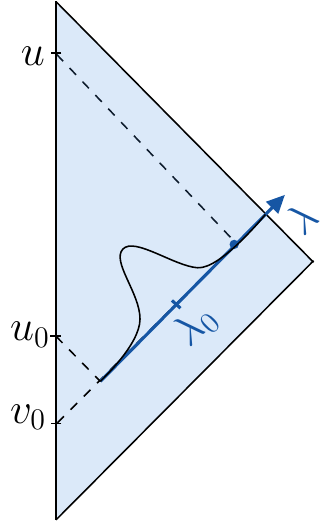}
    \caption{\label{fig:infalling}}
    \end{subfigure}
    \begin{subfigure}[t] {0.48\linewidth}  
\includegraphics[width=0.9\linewidth]{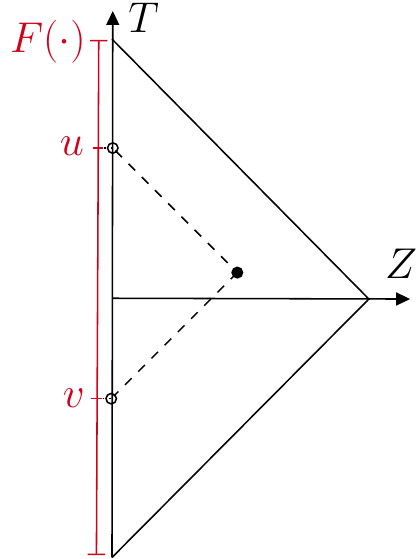}
    \caption{\label{fig:SchwarzianDressing}}
    \end{subfigure}
    \caption{\footnotesize (a) The trajectory of an infalling light ray is parametrized by a boundary time $u$ and its starting position specified by two boundary times $u_0,v_0$. Its affine time $\lambda$ is related to boundary time by $d\lambda=G(u,v_0)\,du$. We show in black a Gaussian detector switching function $\chi(\lambda)$ centered around $\lambda_0$, to be used later around Eq.~\eqref{eq:Proba}. (b) For a given off-shell reparametrization $F(\cdot)$ (in red), a bulk point $(U,V)$ is dressed to the boundary times $u$ and $v$ through the reparametrization mode by $U=F(u),V=F(v)$.}
\end{figure}

In this letter, we argue that non-perturbative quantum effects do indeed lead to drastic changes in the causal structure of black holes. In particular, we show that the necessary condition outlined above for the existence of a black hole interior is \emph{violated} in Jackiw–Teitelboim (JT) gravity~\cite{Jackiw:1984je, Teitelboim:1983ux, Mertens:2022irh}, an exactly solvable toy model of quantum gravity in two spacetime dimensions. 
JT gravity provides a well-understood setting for holographic duality in which the bulk dynamics are entirely encoded in fluctuations of the boundary. Previous works~\cite{Blommaert2019,Mertens:2019bvy, Blommaert:2020yeo, DeVuyst:2022bua, Mertens:2025rpa} have defined and motivated a gravitational dressing analogous to Einstein's radar definition of bulk frames, which acts as a reparameterization of bulk coordinates in terms of two boundary times, $u$ and $v$. In this setup, the metric is promoted to a quantum operator in the boundary Schwarzian theory as $ds^2 =4\,G(u,v)\,du\,dv$, where $G(u,v)$ is the Schwarzian bilocal operator~\cite{Mertens:2017mtv}.\footnote{See \cite{Liu:2024qnh,Tierz:2026iww} for some other recent applications of this procedure.} The affine time taken by an infalling null ray (an idealization of an infinitely accelerated observer), see Fig.~\ref{fig:infalling}, to reach the horizon reads
\begin{equation}
\label{eq:DT}
    \Delta \lambda = - \int_{u_0}^{\infty} du ~\langle G(u,v_0)\rangle.
\end{equation}
Notably, this quantity is defined entirely in terms of the boundary data. The brackets around $G(u,v_0)$ correspond to the expectation value of $G(u,v_0)$ in full-fledged quantum gravity. We will show that non-perturbative gravitational effects render $\Delta \lambda$ infinite, thereby preventing any observer from ever crossing the horizon in finite time. This challenges the idea that the notion of black hole interior has operational meaning in quantum gravity.

 Another powerful way of probing the structure of quantum black hole horizons is by computing observables related to interactions between an infalling observer and its environment. A useful toy model for the infalling observer is the Unruh-DeWitt detector~\cite{Unruh1976,DeWitt:1980hx}, a localized quantum system linearly coupled to a bulk matter field. It offers a valuable tool for probing the structure of gravitationally coupled matter fields, and for characterizing what an infalling observer would ``see'' apart from its own clock~\cite{Blommaert:2020yeo,Franken:2026bff}.

Building on our definition of gravitationally dressed observables, we compute the excitation probability of a detector turned on for for a short amount of time, as shown in Fig.~\ref{fig:infalling}, and find that it deviates significantly both from the semiclassical and Schwarzian disk predictions when sufficiently close to the horizon. As in our companion work~\cite{Franken:2026bff}, quantum-gravitational effects already present in the Schwarzian theory generate a peak in the excitation probability. Once the affine time operator exits the regime of validity of the Schwarzian theory, non-perturbative corrections produce a ramp-plateau structure: the excitation probability decreases from this peak and asymptotes, at infinite affine time, to a constant value that remains above the semiclassical prediction. 

Together, we see our results as a concrete manifestation of the fact that the causal structure of black holes in quantum gravity can drastically deviate from semiclassical expectations, even when the curvature near the horizon is not strong. \\

\textbf{\textit{Gravitational dressing in JT gravity--}} The classical action of JT gravity on a manifold $M$ is given by~\cite{Jackiw:1984je, Teitelboim:1983ux}
\begin{align}
\begin{split}
\label{eq:JTgravityaction}
    S_{\text{JT}}[g, \Phi] &= \dfrac{1}{16\pi G_N}\int_M d^2x\sqrt{-g}\,\Phi(R+2) \\
    &+ \dfrac{1}{8\pi G_N}\int_{\partial M}dt\sqrt{-h}\,\Phi(K-1) + S_0 \chi(M).
\end{split}
\end{align}
Since the dilaton field $\Phi$ appears in the bulk action as a Lagrange multiplier, the theory also admits an equivalent description in terms of the trajectory of the boundary curve $\partial M$ embedded in a rigid AdS$_2$ background. In the limit where the boundary curve $\partial M$ asymptotes to the conformal boundary of AdS$_2$, the trajectory is fully fixed by a single function $F(t)$ called the \emph{Schwarzian mode}, corresponding to the reparametrization from the proper time $t$ of the boundary theory to the Poincar\'e time of the AdS$_2$ bulk. The action for $F(t)$ reads~\cite{Almheiri:2014cka,Jensen:2016pah,Maldacena:2016upp,Engelsoy:2016xyb} 
\begin{equation}
    S[F] = -C\int dt\,\{F, t\},
\end{equation}
where $\{F, t\} \equiv \frac{F'''}{F'} - \frac{3}{2}\left(\frac{F''}{F'}\right)^2$ is the Schwarzian derivative, and $C \propto 1/G_N$ is the Schwarzian coupling constant.

The Schwarzian mode can be used to define a form of gravitational dressing in JT gravity as follows~\cite{Blommaert2019}.
We fix two points along the boundary curve, labeled by boundary times $u$ and $v$ with $u>v$, as shown in Fig.~\ref{fig:SchwarzianDressing}. We then shoot a past-directed null ray at boundary time $u$ and a future-directed null ray from the point at boundary time $v$. The point where these two rays intersect will be labeled by the lightcone coordinates
\begin{equation}
    \label{eq:dressing}
        U=F(u), \qquad
        V=F(v),
\end{equation}
where $U=T+Z$ and $V=T-Z$ and $(T, Z)$ are Poincar\'e coordinates in AdS$_2$. 
The operationally defined metric in terms of the reparametrization function $F$ is then simply
\begin{equation}
\label{eq:metric}
    ds^2=-\dfrac{4F'(u)F'(v)}{(F(u)-F(v))^2}du\,dv.
\end{equation}
This provides an operational definition of a bulk frame in terms of the Schwarzian mode: the radial coordinate in the bulk will simply be given by the difference $z=(u-v)/2$ between the boundary times where the bulk point is anchored, and a timelike coordinate in the bulk can be constructed as $t=(u+v)/2$~\cite{Blommaert2019}. 

An important feature of the bulk radial and time coordinates $t$ and $z$ defined above is that translations in $t$ at fixed $z$ correspond to the flow of a conformal Killing vector field in the bulk, for every off-shell configuration of the Schwarzian mode $F(t)$. In fact, the requirement that boundary time translations be extended into the bulk as a conformal Killing vector essentially fixes this dressing completely~\cite{Mertens:2025rpa}. Since conformal Killing vector fields are essential to geometric modular flow~\cite{Sorce:2024zme}, which underlies much of the recent work on gravitational von Neumann algebras via the modular crossed product~\cite{WittenCrossedProduct, Chandrasekaran:2022cip,  Chandrasekaran:2022eqq, AliAhmad:2023etg, Jensen:2023yxy, Kudler-Flam:2023hkl,  DeVuyst:2024fxc}, this dressing is the only viable candidate for extending that construction in Schwarzian language beyond the semiclassical regime. From this, one can then study various quantum gravitational effects on matter-coupled JT gravity by promoting the Schwarzian mode $F$ to a truly dynamical degree of freedom~\cite{Mertens:2019bvy, Blommaert:2020yeo, DeVuyst:2022bua, Mertens:2025rpa}.

A black hole at inverse temperature $\beta$ is described by parametrizing the Schwarzian mode as $F(t)=\frac{\beta}{\pi}\tanh\left(\frac{\pi}{\beta}f(t)\right)$. The metric~\eqref{eq:metric} then becomes
\begin{align}
\begin{split}\label{eq:metricSchwarzianBilocal}
    ds^2 &= -\dfrac{4\pi^2}{\beta^2}\dfrac{f'(u)f'(v)}{\sinh^2\left(\frac{\pi}{\beta}(f(u)-f(v))\right)}\, du\,dv \\
    &\equiv G(u, v)\, 4\,du\,dv,
\end{split}
\end{align}
where $G(u,v)$ denotes the thermal Schwarzian bilocal. Once we promote $f(v)$ to a dynamical variable, $G(u, v)$ becomes an operator in the quantum gravity theory. In particular, the average of its two time orderings is a Hermitian operator~\cite{Blommaert2019}\footnote{We leave the averaging over time orderings implicit in the following.} whose semiclassical limit leads precisely to the metric of the black hole patch in AdS$_2$. More generally, its expectation value in a thermal state can be computed by a gravitational path integral, which for a general operator $\mathcal{O}$ in the Schwarzian theory is
\begin{equation}\label{eq:schwarzianexpval}
    \langle{\mathcal{O}}\rangle_\beta = \dfrac{1}{Z(\beta)}\int [\mathcal{D}f]\,\mathcal{O}[f] e^{C\int_0^\beta d\tau \left\{\tan \frac{\pi}{\beta}f,\tau\right\}},
\end{equation}
where $Z(\beta)$ is the Schwarzian partition function~\cite{Mertens:2022irh}. The integral over $f$ ranges over all non-equivalent configurations with $f'(\tau)>0$ satisfying the periodicity constraint $f(\tau+\beta)=f(\tau)+\beta$, characterizing the thermofield double state at inverse temperature $\beta$. When $\mathcal{O}$ is the thermal Schwarzian bilocal defined in Eq.~\eqref{eq:metricSchwarzianBilocal}, its expectation value~\eqref{eq:schwarzianexpval} is known exactly~\cite{Mertens:2017mtv,Bagrets:2017pwq,Blommaert:2018oro,Yang:2018gdb,Iliesiu:2019xuh}. The expectation value of the metric~\eqref{eq:metricSchwarzianBilocal} then becomes
\begin{equation}\label{eq:effectivemetricschwarzian}
    \langle ds^2\rangle= \mathcal{G}_\beta(z)\, 4\,du\, dv,
\end{equation}
where $z=(u-v)/2$ and
\begin{equation}
    \mathcal{G}_\beta(z)\equiv \langle G(u, v)\rangle_\beta.
\end{equation}
The exact result for $\mathcal{G}_\beta(t)$ in the Schwarzian theory is given by \cite{Mertens:2017mtv}
\begin{equation}\label{eq:Schwarziantwoptfunction}
    \mathcal{G}_\beta(z)\propto \int_{\mathbb{R}_+^2} dM dE e^{-\beta M}\rho_0(M)\rho_0(E)e^{-2i(E-M)z}\abs{\mathcal{O}_{ME}^1}
^2,\end{equation}
where $\rho_0(E)$ denotes the Schwarzian density of states, and $\mathcal{O}_{ME}^1$ is the matrix element of a conformal primary with weight $\Delta=1$ in a $(0+1)$-dimensional CFT. Their exact form is given in Eq.~\eqref{eq:density} of the Appendix. This object is a boundary-intrinsic quantity that gives rise to an emergent bulk geometry via~\eqref{eq:effectivemetricschwarzian} due to the gravitational dressing of bulk points to the Schwarzian boundary. \\

\textbf{\textit{Non-perturbative effects push the horizon infinitely far away--}} The metric~\eqref{eq:effectivemetricschwarzian} includes all the Schwarzian corrections to the bulk physics, since $\mathcal{G}_\beta(z)$ is known exactly in the Schwarzian theory. This, however, is not the end of the story in JT gravity since wormhole effects have to be taken into account, weighted in the action~\eqref{eq:JTgravityaction} by the topological term. The full action we should consider is
\begin{equation}
    S_{\text{JT}} = -C\int dt\,\{F, t\} +S_0 \chi(M),
\end{equation}
where $\chi(M)$ is the Euler characteristic of the bulk manifold $M$. When JT gravity is viewed as a dimensional reduction of a higher-dimensional theory, $S_0\sim1/G_N$ accounts for the extremal black hole entropy. In the gravitational path integral, higher topologies are suppressed as $S_0\to\infty$, but whenever $S_0$ is finite, they have to be included. We can roughly think of $e^{S_0}$ as playing the role of $N$, counting a discrete number of degrees of freedom in our quantum system.

It is now well-known that including higher topology effects makes JT gravity behave like a random matrix integral~\cite{Saad:2019lba}. Random matrix theory provides a universal description of systems displaying chaotic behavior, where expectation values can be obtained by treating the system's Hamiltonian as a random variable over some statistical ensemble. This has been observed explicitly in examples such as the SYK model~\cite{SYK1,SYK2, Cotler:2016fpe}, and is also expected to be true for microscopic descriptions of black holes in quantum gravity more generally. 

Treating JT gravity as a strongly chaotic system obeying random matrix statistics has direct consequences to the effective metric observable~\eqref{eq:effectivemetricschwarzian}. In particular, if we think of the two-point function that defines the effective metric as an observable in a statistical ensemble, then $\mathcal{G}_\beta(t)$ in Eq.~\eqref{eq:Schwarziantwoptfunction} should effectively be replaced with
\begin{equation}
\label{eq:matrixG}
    \overline{\mathcal{G}}_\beta(z) \propto \int_{\mathbb{R}_+^2} dM dE \,e^{-\beta M} \overline{\rho(M)\rho(E)}e^{-2i(E-M)z}\abs{\mathcal{O}_{ME}^1}^2,
\end{equation}
where $\overline{\rho(M)\rho(E)}$ is the pair density correlator of the random matrix ensemble, see the Appendix. 

\begin{figure}
    \includegraphics[width=0.8\linewidth]{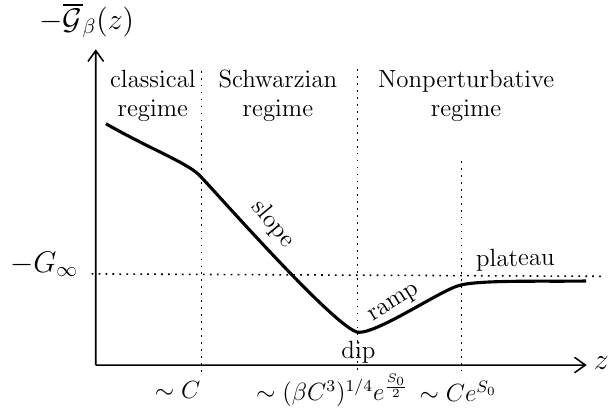}
    \caption{\footnotesize Averaged value of the metric operator as a function of $2z=u-v$, in a log-log plot. }
    \label{fig:plateau}
\end{figure}

The qualitative behavior of $\overline{\mathcal{G}}_\beta(z)$ is shown in Fig.~\ref{fig:plateau} for the case of a very quantum black hole, $\beta \gg C$. The asymptotic AdS boundary is at $u=v$, or $z=0$ in the argument of $\overline{\mathcal{G}}_\beta$. The bilocal displays initial phases of decay, first in the classical regime (where the Schwarzian $F(t)$ is well-approximated by its saddlepoint value $F(t)\propto\tanh(\pi t/\beta)$ and the metric \eqref{eq:metric} reduces to the asymptotic AdS$_2$ metric), and then in the Schwarzian regime, $z\gtrsim C$, where fluctuations of the Schwarzian mode become relevant. After exponentially long times, however, this decay reaches a minimum value (the dip), followed by a phase of linear growth (the ramp); finally, for $z\gtrsim Ce^{S_0}$, the two-point function reaches a plateau with chaotic fluctuations averaging to a finite value $-G_{\infty}$ \cite{Saad:2018bqo, Blommaert2019, Saad:2019pqd}. The existence of the ramp and plateau regions is due to the replacement of the product of the densities of states $\rho_0(M)\rho_0(E)$ by the pair density correlator $\overline{\rho(M)\rho(E)}$, and therefore encodes information about the microscopic statistics of the energy eigenvalues that is not available semi-classically.

In conclusion, at very late boundary times $u-v\sim C e^{S_0}$, corresponding to the near-horizon region in the semiclassical geometry, the effective metric becomes flat,
\begin{equation}
\label{eq:plateaumetric}
    \langle ds^2\rangle = G_{\infty}\, 4\,du\,dv.
\end{equation}
 The affine time $\Delta \lambda$ defined in Eq. \eqref{eq:DT} therefore grows linearly in $u$ at very late time, and diverges as we send $u\to\infty$, as shown in Fig.~\ref{fig:lambda} of the Appendix.

 Similarly, the proper time of a timelike observer approaching the would-be horizon at $u\to\infty$ is also infinite, since the metric at very late times is just flat. The causal structure near the horizon becomes identical to an asymptotically flat spacetime, and the would-be horizon is analogous to null infinity $\mathscr{I}^\pm$. It is in this sense that the region behind the black hole horizon becomes fundamentally inaccessible to an observer from outside, even if they try to jump in. 
 
 An alternative but related test is to generalize the classical definition of the horizon as the null hypersurface where $G(u,v)=0$. The late time behavior of $\bar{\mathcal{G}}_{\beta}(u,v)$ shows that such a surface does not exist in JT gravity in an averaged sense.

Eq.~\eqref{eq:plateaumetric} is calculated as an ensemble average. An observer falling instead towards a black hole microstate (modeled as a single draw of the matrix ensemble) will experience large fluctuations in its affine time, due to the erratic nature of the plateau region. However, the conclusion that $\Delta\lambda$ diverges remains valid even when considering such a specific microstate geometry.
\\

\textbf{\textit{Infalling Unruh-DeWitt detector--}} Now we describe the infalling observer's experience more quantitatively. We model the observer as a local detector linearly coupled to a bulk massless scalar field $\phi$. Specifically, we consider an Unruh-DeWitt detector~\cite{Unruh1976,DeWitt:1980hx}, whose interaction action reads
\begin{equation}\label{eq:generalUDW}
    S_I = q\int\,d\lambda\,\chi(\lambda)\mu(\lambda) k^{\mu}\nabla_{\mu}\phi(\mathrm{x}(\lambda)),
\end{equation}
where $q$ is a coupling constant, $\lambda$ is the time along the trajectory, $\mu$ is a monopole operator on the detector Hilbert space, $\mathrm{x}(\lambda)$ is the trajectory of the detector, and $k^{\mu}$ is the vector tangent to the trajectory, $k^{\mu}=dx^{\mu}/d\lambda$. The switching function $\chi(\lambda)$ governs the time-dependence of the interaction's strength. To connect with the previous section, we chose $\mathrm{x}(\lambda)$ to be a null infalling trajectory and $\lambda$ its affine time.

The simplest observable that can be computed is the probability that the detector is excited from a ground state $\ket{g}$ to an excited state $\ket{e}$. At leading order in a series expansion in the coupling constant $q$, this probability can be written as
\begin{align}
\begin{split}\label{eq:excprobgeneral2}
    P_{\text{exc}} &= q^2 \int d\lambda d\lambda' \chi(\lambda) \chi(\lambda')e^{-i\omega(\lambda-\lambda')} \partial_{\lambda}\partial_{\lambda'} W(\lambda,\lambda').  
\end{split}
\end{align}
In the equation above, we introduced $\omega$ as the internal energy gap of the detector between the ground and excited states in its proper frame, and $W(\lambda,\lambda') = \langle \phi(\mathrm{x}(\lambda))\phi(\mathrm{x}(\lambda'))\rangle$ the two point-function along the trajectory. We dropped the detector matrix element $\vert\bra{e}\mu(0)\ket{g}\vert^2$ for simplicity. 

Quantum-gravitational corrections to the excitation probability appear from the gravitational dressing of affine time and of the two-point function. The gravitationally dressed affine time has already been shown to be promoted to an operator via Eq.~\eqref{eq:metricSchwarzianBilocal}. Similarly, as shown in~\cite{Blommaert2019,Blommaert:2020yeo, Franken:2026bff}, applying the dressing~\eqref{eq:dressing} to the two-point function of a free massless scalar yields
\begin{equation}\label{eq:dressedWightmanLL}
    W(\mathrm x, \mathrm x') = \dfrac{1}{4\pi}\int_v^u dt \int_{v'}^{u'}dt' \,G(t,t').
\end{equation}
Putting everything together, we have
\begin{equation}
\label{eq:Proba}
    P_{\rm exc} = \frac{q^2}{4\pi} \hspace{-1mm}\int \hspace{-1mm} du  du' \chi(\lambda(u))\chi(\lambda(u'))e^{-i\omega\int_{u'}^u dt\, G(t,v) }G(u,u').
\end{equation}
To restrict the coupling to a localized region, we choose a Gaussian switching function centered around $\lambda_0$ with a characteristic duration of the interaction in affine time $\sigma$ (see Fig.~\ref{fig:infalling}), $
    \chi(\lambda) = e^{-\frac{(\lambda-\lambda_0)^2}{2\sigma^2}}$. We then treat~\eqref{eq:Proba} as an observable in the full quantum gravity theory, and compute its expectation value in an approximation that is suitable for the near-horizon limit of very large $u$; for more details, see~\cite{Franken:2026bff} and Appendix. 

\begin{figure}[ht]
    \includegraphics[width=0.95\linewidth]{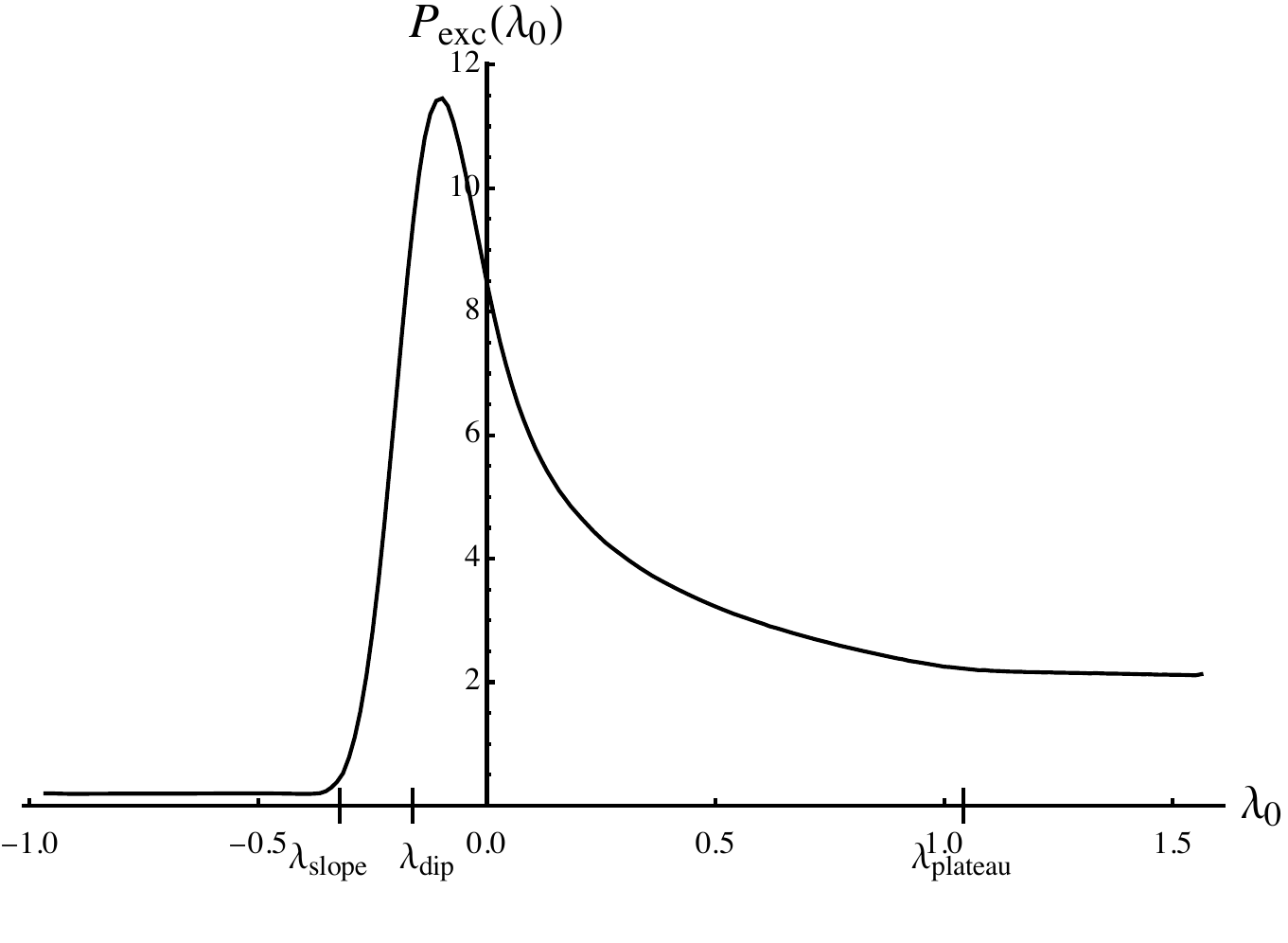}
    \caption{\footnotesize Excitation probability (in units of $q^2$) of a detector in the near-gapless regime ($\sigma \omega \ll 1$) for a Gaussian switching function centered around $\lambda_0$, as a function of $\lambda_0$. We used $\sigma \omega=0.1, C=1,\beta=15,S_0=5$ for the numerical plot.}
    \label{fig:Pexc}
\end{figure}

We compute $P_{\rm exc}$, taking the full non-perturbative expression of $W(\mathrm{x},\mathrm{x}')$, and approximating $\lambda-\lambda'$ by its dominant behavior in the different scaling regimes shown in Fig.~\ref{fig:plateau}. We use $\sigma \ll 1/C$ so that the detector can distinguish between the different regimes, and gapless limit $\sigma\omega\ll 1$ to simplify numerical computations. The resulting non-perturbative excitation probability is shown in Fig.~\ref{fig:Pexc}.
The different regimes correspond in the radial $z$-coordinate to $z_{\text{slope}}\sim C$, $z_{\text{dip}} \sim (\beta C^3)^{1/4}e^{S_0/2}$, and  $z_{\text{plateau}}\sim Ce^{S_0}$.

\begin{figure}
    \begin{subfigure}[t] {0.48\linewidth}  
\includegraphics[width=1\linewidth]{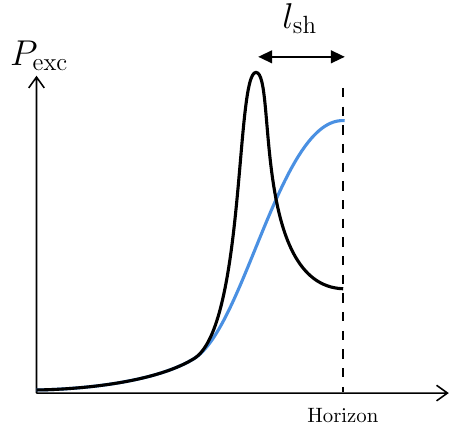}
\caption{\footnotesize Non-perturbative shift of the horizon}
\label{fig:peakcompare}
    \end{subfigure}
    \begin{subfigure}[t] {0.48\linewidth}  
\includegraphics[width=0.8\linewidth]{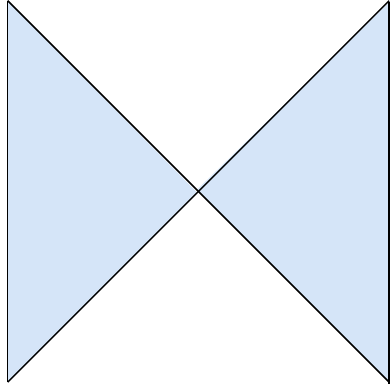}
\caption{\footnotesize Quantum gravity Penrose diagram}
\label{fig:Penrose}
    \end{subfigure}    \hfill
    \caption{\footnotesize  (a) Excitation probability of an infalling detector at the non-perturbative level (in black) and in the Schwarzian theory (blue). While the Schwarzian curve is peaked at the horizon, the non-perturbative result shows a stretched horizon very close to the classical horizon, followed by a plateau in the $\mathscr{I}^\pm$-like region. (b) Penrose diagram of an eternal black hole. The causal structure of quantum spacetime, defined as $\langle ds^2\rangle= \overline{\mathcal{G}}_\beta(z)\, 4\,du\, dv$, does not contain any interior region, as the infalling null rays never reach the horizon---much like $\mathscr{I}^\pm$ in asymptotically flat spacetimes.}
\end{figure}

The initial constant value is consistent with the prediction from QFT in a fixed AdS$_2$ background, which was reviewed in~\cite{Franken:2026bff}. The increase in excitation during the slope regime has been studied extensively in our companion work~\cite{Franken:2026bff}, and the height of the peak decreases faster than any power law in $\omega$, which we argued is a sufficient condition for the absence of a firewall. The non-perturbative corrections on $\lambda$ lead to a shift of the peak in excitation probability from the classical horizon, which is now located at $\lambda\rightarrow \infty$, to an affine parameter $\lambda\sim e^{\# S_0}$. In this sense, if an infalling observer uses the peak of the excitation probability as a diagnostic for the horizon as in \cite{Franken:2026bff}, he would measure a ``stretched horizon" at a proper distance $l_{\rm sh}$ that is doubly-exponential in $S_0$ to the classical horizon. After the peak, starting at a point with proper distance to the classical horizon $\propto e^{-2\pi \frac{C}{\beta}e^{S_0}}$, the excitation probability asymptotes to a constant that is distinct from the semiclassical value, reflecting the non-perturbative corrections to the black hole horizon structure very near the horizon. We sketch the qualitative difference between the Schwarzian and wormhole calculations in Fig.~\ref{fig:peakcompare}.\\

\textbf{\textit{Conclusion--}} In this letter, we studied a manifestation of strong quantum-gravitational effects in the experience of local observers in the vicinity of black hole horizons. We did so by presenting a concrete realization of gravitational dressing of local observers in JT gravity, where quantum effects beyond the semiclassical regime are analytically tractable. The response of an infalling probe in this setup is summarized by the plot in Figure~\ref{fig:Pexc}, which shows that a detector coupled to a gravitationally dressed quantum field for a finite amount of affine time is sensitive to quantum-gravitational effects corresponding to different phases of its approach to the horizon.

Our main qualitative conclusion is schematically pictured in the ``quantum gravity Penrose diagram'' of Fig.~\ref{fig:Penrose}, where the interior region is fully blocked off. The fact that localized observers are unable to reach the horizon ultimately implies that the black hole interior does not exist in non-perturbative quantum gravity in an operational sense. This adds a further wrinkle to the beyond-the-horizon firewall scenario of \cite{Stanford:2022fdt,Iliesiu:2024cnh,Blommaert:2024ftn}: when insisting on gravitationally dressed observables, the infalling probe does not even reach the horizon to begin with.

We believe that the lessons learned in this work may extend well beyond two-dimensional quantum gravity, as they ultimately rely on the chaotic nature of the black hole spectrum, which is a universal property of quantum black holes~\cite{Cotler:2016fpe}. In particular, the qualitative difference between the results obtained with finite vs. infinite $S_0$ can be seen as an operational realization in JT gravity of the result in AdS/CFT~\cite{Witten:2021unn,Leutheusser:2021qhd, Leutheusser:2021frk} linking the existence of a connected black hole interior and sharp causal horizon structure in the bulk to the transition from a type I to a type III$_1$ von Neumann algebra for the boundary CFT in the limit $N\rightarrow\infty$. One may also interpret the inaccessibility of the black hole interior as an operational manifestation of bulk factorization as articulated in \cite{Boruch:2024kvv,Balasubramanian:2024yxk}.

\textbf{\textit{Acknowledgments--}} We thank Marine De Clerck, Jackson Fliss, Kyriakos Papadodimas, Simon Ross, Arvin Shahbazi-Moghaddam, and Thomas Tappeiner for insightful discussions. We acknowledge financial support from the European Research Council (grant BHHQG-101040024). Funded by the European Union. Views and opinions expressed are however those of the authors only and do not necessarily reflect those of the European Union or the European Research Council. Neither the European Union nor the granting authority can be held responsible for them.

\appendix

\section{Appendix: Computation of the excitation probability}
\label{app:computation}

We now compute the excitation probability~\eqref{eq:Proba} in the different regimes shown in Fig.~\ref{fig:plateau}. The different regimes of interest for $\overline{\mathcal G}_{\beta}(t)$ are
\begin{enumerate}
    \item Classical regime:\footnote{For ease of the technical calculation, we take here $C\ll \beta$, where the semiclassical approximation reduces to a power law. This will not impact the other regions, which is our main focus of this letter.} \\
    $\overline{\mathcal{G}}_{\beta}(t) \simeq - A_{\rm cl}\dfrac{\pi^2}{\beta^2 \sinh^2\left(\frac{\pi}{\beta}t\right)},$ with $t \lesssim C$.
    \item `Slope'' - Schwarzian (disk) regime:\\$\overline{\mathcal{G}}_{\beta}(t) \simeq - A_{\rm Schw}t^{-3}$ with $C \lesssim t\lesssim (2\beta C^3)^{1/4}e^{S_0/2}$.
    \item ``Ramp'' - Non-perturbative regime: \\ $\overline{\mathcal{G}}_{\beta}(t) \simeq  - A_{\rm ramp}t$ with $(2\beta C^3)^{1/4}e^{S_0/2} \lesssim t\lesssim Ce^{S_0}$.
    \item ``Plateau'' - Doubly non-perturbative regime:\\ $\overline{\mathcal{G}}_{\beta}(t) \simeq - G_{\infty}$ with $Ce^{S_0} \lesssim t $.
\end{enumerate}
The exact value of constants $A_{\rm cl}, A_{\rm disk}, A_{\rm ramp}$ and $G_{\infty}$ can be calculated explicitly from 
\begin{equation}\label{eq:exactbilocal}
    \overline{\mathcal{G}}_\beta(t)=\frac{e^{S_0}}{Z(\beta)} \int_{\mathbb{R}_+^2} dM dE e^{-\beta M}\overline{\rho(M)\rho(E)}e^{-i(E-M)t}\abs{\mathcal{O}_{ME}^1}^2,
\end{equation}
and using the known pair density correlator~\cite{mehta2004random}
\begin{align}\label{eq:pairdensitycorrelator}
\begin{split}
    \overline{\rho(M)\rho(E)} &\simeq \rho_0(E)\rho_0(M) - \frac{1}{2\pi^2(E-M)^2}\\
    &+ \frac{\cos\left(2\pi\rho_0(M)(E-M)\right)}{2\pi^2(E-M)^2} + \rho_0(E)\delta(E-M),
\end{split}
\end{align}
where
\begin{align}
\begin{split}
\label{eq:density}
    &\rho_0(E) = e^{S_0}\dfrac{C}{2\pi^2}\sinh\left(2\pi\sqrt{2CE}\right), \\
    &|\mathcal{O}^{1}_{ME}|^2 \equiv 2e^{-2S_0}\dfrac{\Gamma(1\pm i\sqrt{2CM}\pm i\sqrt{2CE})}{(2C)^2}, \\
    &Z(\beta) = \dfrac{1}{4\pi^2}\left(\dfrac{2\pi C}{\beta}\right)^{3/2}e^{S_0 + \frac{2\pi^2 C}{\beta}} \equiv e^{S_0}Z_0(\beta).
\end{split}
\end{align}
The approximation above for $\overline{\rho(M)\rho(E)}$ is appropriate in the limit of small energy differences $\abs{E-M} \ll 1/C$ and away from the edge of the spectrum at $E, M=0$. This is the region that dominates the integral in Eq.~\eqref{eq:exactbilocal} in the limit of large $t$. The first term in Eq.~\eqref{eq:pairdensitycorrelator} is the factorized contribution from the two disconnected disks, while the remaining terms are non-perturbative in $1/S_0$ and come from wormhole corrections and matrix statistics, that give characteristic high frequency wiggles with spacing $\sim e^{-S_0}$. The classical and Schwarzian disk regimes is dominated by the early and late time behaviors of the first term, respectively. The ramp appears when including the second term. Finally, the plateau comes from the third term which cancels the linear growth of the ramp, and the value of $G_{\infty}$ comes from the contribution of the fourth (plateau) term. We have:
\begin{gather}
\label{eq:constants}
    A_{\rm cl}=1, \qquad\qquad\quad\,\,
    A_{\rm Schw} = \frac{C}{4\pi Z_0(\beta)},\\
    A_{\rm ramp} = \frac{e^{-2S_0}}{2\pi\beta Z_0(\beta)(2C)^2}, \qquad
    G_{\infty} = \frac{e^{-S_0}}{2C^2}.
\end{gather}
To compute the detector response in more detail in this setup, we will approximate the relation between boundary time $t$ and affine time $\lambda$ to be given by a piecewise defined function defined by the various regimes of $\overline{\mathcal{G}}_\beta(t)$ outlined earlier, with the transitions occurring sharply at the times $t_{\rm slope}$, $t_{\rm dip}$, and $t_{\rm plateau}$. In other words, we will take
\begin{align}\label{eq:piecewiseaffinetime}
    \dfrac{d\lambda}{d t} &= \dfrac{A_{\rm cl}}{t^2}\Theta(t_{\rm slope}-t) \nonumber \\
    &+ \dfrac{A_{\rm Schw}}{t^3}\Theta(t-t_{\rm slope})\Theta(t_{\rm dip}-t) \nonumber \\
    &+ A_{\rm ramp} t \Theta(t-t_{\rm dip}) \Theta(t_{\rm plateau}- t) \nonumber \\
    &+ G_\infty \Theta(t-t_{\rm plateau}),
\end{align}
where we have abbreviated $u-v\equiv t$. If we now demand continuity of the bilocal $\overline{\mathcal{G}}_{\beta}(t)$ across the different regimes and use the results summarized in Eq.~\eqref{eq:constants}, the transition times between these regimes are given quantitatively by
\begin{align}
    t_{\rm slope} &= \frac{C}{4\pi Z_0(\beta)},\\
    t_{\rm dip}&= (2\beta C^3)^{1/4}e^{S_0/2},\\
    t_{\rm plateau}&=4\pi\beta Z_0(\beta)e^{S_0}.
\end{align}
The ``dip time'' $t_{\rm dip}$ at which the slope and ramp meet is theory-dependent. On the other hand, the scalings and other time-scales are universal for random matrix theory and therefore for quantum black holes~\cite{Cotler:2016fpe,Dyer:2016pou,Gharibyan:2018jrp}. 

\begin{figure}[ht]
    \includegraphics[width=1\linewidth]{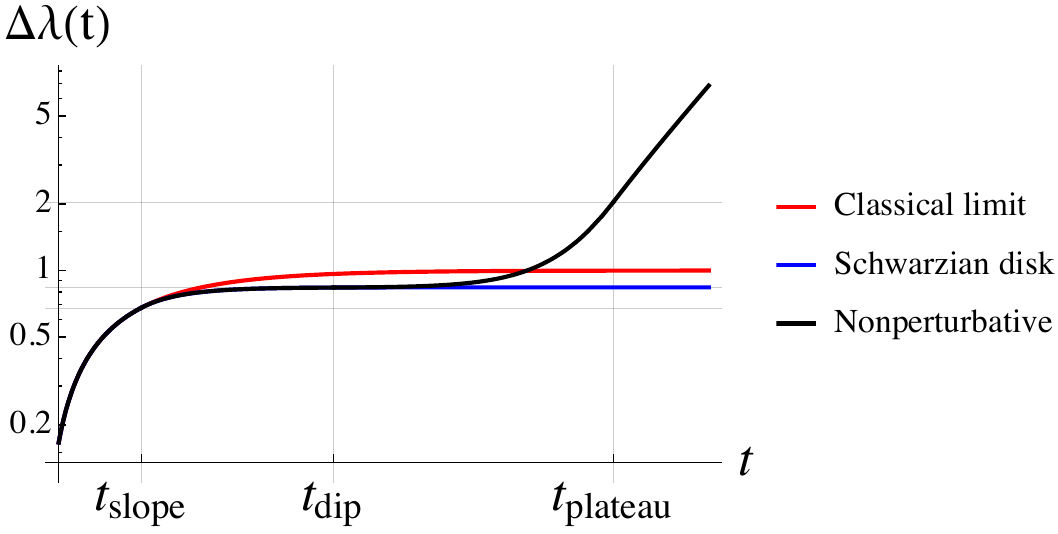}
    \caption{\footnotesize Log-Log plot of the affine time difference between an infalling observer and a fixed point at $u-v\leq t_{\rm slope}$, as a function of $t=u-v$, at different level of approximation. While $\Delta\lambda$ goes to a constant in the classical limit and in the Schwarzian disk theory, it diverges when including higher-topology effects. We used $C=1,\beta=15,S_0=5$ for the numerical plot.}
    \label{fig:lambda}
\end{figure}

From this, and assuming that we start the infalling trajectory at a point close to the boundary $(u_i=v+\epsilon,v)$ with $\epsilon\ll 1$, we find that the infalling observer is in the classical regime for an affine time $\Delta\lambda_{\rm classical}\sim 1/\epsilon$. It is in the Schwarzian disk regime for $\Delta\lambda_{\rm Schw}\sim 1/C$, in the ramp regime for $\Delta\lambda_{\rm ramp}\sim \beta/C^2$, and on the plateau for an infinite amount of time.
Integrating~\eqref{eq:piecewiseaffinetime} and setting $\lambda=-\infty$ at $t=0$, we have
\begin{widetext}
\begin{equation}\label{eq:lambdaoftgeneral}
    \lambda(t)= \left\{ \begin{array}{cc}
      &-\dfrac{A_{\rm cl}}{t}, \quad t<t_{\rm slope},   \\
      &\dfrac{A_{\rm Schw}}{2}\left(\dfrac{1}{t_{\rm slope}^2} - \dfrac{1}{t^2}\right) - \dfrac{A_{\rm cl}}{t_{\rm slope}},   \quad t_{\rm slope}< t< t_{\rm dip},\\
      &\dfrac{A_{\rm ramp}}{2}(t^2-t_{\rm dip}^2) +\dfrac{A_{\rm Schw}}{2}\left(\dfrac{1}{t_{\rm slope}^2} - \dfrac{1}{t^2_{\rm dip}}\right) - \dfrac{A_{\rm cl}}{t_{\rm slope}}, \quad t_{\rm dip}< t< t_{\rm plateau}, \\
      &G_{\infty}(t-t_{\rm plateau}) + \dfrac{A_{\rm ramp}}{2}(t_{\rm plateau}^2-t_{\rm dip}^2) +\dfrac{A_{\rm Schw}}{2}\left(\dfrac{1}{t_{\rm slope}^2} - \dfrac{1}{t^2_{\rm dip}}\right) - \dfrac{A_{\rm cl}}{t_{\rm slope}} \quad t>t_{\rm plateau}.
    \end{array}\right.
\end{equation}
\end{widetext}
The function $\lambda(t)$ is plotted in Fig.~\ref{fig:lambda}.

The formula for the excitation probability can be put in the general form
\begin{align}\label{eq:excprobgeneralRMT}
    P_{\text{exc}}= \dfrac{q^2}{4\pi Z_0(\beta)}\int dE\,dM\,e^{-\beta M}\overline{\rho(E)\rho(M)}&\abs{\mathcal{O}^{1}_{ME}}^2 \\
    \times&\abs{\mathcal{I}_\omega(E-M)}^2,\nonumber
\end{align}
where we define
\begin{equation}
    \mathcal{I}_\omega(\nu)=\int_0^\infty du\,\chi(\lambda(u))e^{-i\omega\lambda(u)}e^{-i\nu u}.
\end{equation}
This applies throughout the detector's entire history. The integrals of Eq.~\eqref{eq:excprobgeneralRMT} can be computed numerically by using the explicit form of $\lambda(t)$ given by Eq.~\eqref{eq:lambdaoftgeneral}. The resulting excitation probability as a function of $\lambda_0$ (the affine time where the Gaussian is peaked) is shown in Fig.~\ref{fig:Pexc}.

We can make analytical progress evaluating the excitation in the plateau region $t>t_{\text{plateau}}$, where $\lambda(u)= G_\infty u + \Lambda_0$. Here, $\Lambda_0$ is just an abbreviation for all the constant terms appearing in the plateau regime of $\lambda(t)$ in Eq.~\eqref{eq:lambdaoftgeneral}. In this case, we have
\begin{align}
    \mathcal{I}_\omega(\nu)= \left(e^{i \#}\right) \dfrac{\sigma}{G_\infty}\sqrt{2\pi} \,e^{-\frac{1}{2}\left(\sigma\omega + \frac{\sigma \nu}{G_\infty}\right)^2},
\end{align}
where the prefactor $\left(e^{i \#}\right)$ is a pure phase that depends on the parameters $\Lambda_0$ and $\lambda_0$. Upon taking the modulus squared, we find
\begin{equation}\label{eq:IintegralGaussian}
    \abs{\mathcal{I}_\omega(\nu)}^2 = \dfrac{2\pi \sigma^2}{G_\infty^2} e^{-\left(\sigma\omega + \frac{\sigma \nu}{G_\infty}\right)^2}.
\end{equation}
We note that this does not depend on $\lambda_0$ (the affine time where the Gaussian is centered), automatically implying that the excitation probability at very late times is invariant under time translations of the switching function. This is just like the limit of our setup to a QFT in curved spacetime, reviewed e.g. in \cite{Franken:2026bff}.

We also immediately see that $\abs{\mathcal{I}_\omega(\nu)}^2$ in this regime is a Gaussian peaked at $\nu=-G_{\infty}\omega$ with characteristic width $G_\infty/\sigma$. Since $G_\infty \propto e^{-S_0}$, we learn that both the location of the peak and its spread in $\nu$ will be very small as long as we assume that $\sigma$ is not non-perturbatively small and $\omega$ is not exponentially large in $S_0$. Changing integration variables from $(M, E)$ to $(M, \nu\equiv E-M)$ in Eq.~\eqref{eq:excprobgeneralRMT} and inserting Eq.~\eqref{eq:pairdensitycorrelator} for the pair density correlator in the limit of small energy separations,
we get
\begin{align}
\label{eq:Ppairdensity}
    P_{\text{exc}}= \dfrac{q^2}{4\pi Z_0(\beta)}&\int d\nu\,dM\,e^{-\beta M}\abs{\mathcal{I}_\omega(\nu)}^2\abs{\mathcal{O}^{1}_{M, M+\nu}}^2\nonumber \\
    &\times\Big(\rho_0(M)\rho_0(M+\nu) \nonumber \\
    &- \dfrac{\sin^2\left[\pi \rho_0(M)\nu\right]}{\pi^2 \nu^2}+\rho_0(M)\delta(\nu)\Big).
\end{align}
The most important qualitative effect from the inclusion of higher topologies is the addition of the sine kernel to the pair density correlator, which leads to \emph{level repulsion} of the black hole system when $E \approx M$ (or $\nu \approx 0$). This manifests as a drastic reduction of the excitation and emissions probabilities of a detector in the limit of very low frequencies $\nu \sim e^{-S_0}$.

Solving the integral in $\nu$, the total excitation probability for a detector coupled to black hole at inverse temperature $\beta$ can thus be written as
\begin{equation}\label{eq:probcanonicalvsmicrocanonical}
    P_{\text{exc}}(\beta) = \dfrac{1}{Z(\beta)}\int_0^\infty dM\,e^{-\beta M}\rho_0(M)P_{\text{exc}}(M),
\end{equation}
where the explicit form of $P_{\text{exc}}(M)$ is given by
\begin{align}\label{eq:excprobmicrocanonical}
    P_{\text{exc}}(M) = e^{S_0} q^2\sigma C  \sqrt{\dfrac{CM}{2\pi}}.
\end{align}
We interpret $P_{\text{exc}}(M)$ as the excitation probability of a detector falling into a black hole in the microcanonical ensemble at fixed energy $M$, whereas $P_{\text{exc}}(\beta)$ is the excitation probability of a detector falling into a black hole in the canonical ensemble at fixed inverse temperature $\beta$. The final integral in Eq.~\eqref{eq:probcanonicalvsmicrocanonical} can also be computed in closed form, and gives
\begin{align}\label{eq:analyticalcanonical}
    P_{\text{exc}}(\beta) = e^{S_0}q^2\sigma C&\Bigg(\sqrt{\dfrac{C}{2\pi^2\beta}}e^{-\frac{2\pi^2 C}{\beta}} \\
    &+\sqrt{\pi } \left( \frac{C}{\beta}+\frac{1}{4\pi^2}\right) \erf\left(\sqrt{\frac{2\pi^2 C}{\beta}}\right)\Bigg).\nonumber
\end{align}
In the limit of $\beta\ll C$, corresponding to a large black hole, the leading-order contribution to this is simply given by
\begin{equation}
    P_{\text{exc}}(\beta\ll C) \simeq  e^{S_0}q^2\sigma C\sqrt{\pi}\frac{C}{ \beta},
\end{equation}
which is precisely what we get if we evaluate $P_{\text{exc}}(M)$ from Eq.~\eqref{eq:excprobmicrocanonical} at $M=M_{\text{th}} = 2\pi^2 C/\beta^2$, the thermal ADM energy of a black hole at inverse temperature $\beta$ in classical JT gravity. Away from this regime, however, the microcanonical and canonical ensembles differ significantly, so the full answer in Eq.~\eqref{eq:analyticalcanonical} is more appropriate.

Note that in the infinite $\sigma$ limit, the plateau contribution to the excitation probability \eqref{eq:Ppairdensity} contracts to a contribution with delta support at $\omega =0$ \cite{Blommaert:2020yeo}.

\bibliographystyle{apsrev4-1}
\bibliography{references}

\end{document}